**PAPER • OPEN ACCESS**

# Temperature measurement methods in an experimental setup during bone drilling: A brief review on the comparison of thermocouple and infrared thermography

To cite this article: Md Ashequl Islam *et al* 2021 *J. Phys.: Conf. Ser.* **2129** 012096

View the article online for updates and enhancements.

**You may also like**

- Scaling of divertor power footprint width in RF-heated type-III ELMy H-mode on the EAST superconducting tokamak
  L. Wang, H.Y. Guo, G.S. Xu et al.

- Infrared cameras as accessories to smartphones: facts you need to know
  Michael Vollmer and Klaus-Peter Möllmann

- Installation of IR Camera in Smart Watches for Facial Expression Recognition
  K. Sathish, Shaikh Mojamill, G. Sai Priya et al.

# Temperature measurement methods in an experimental setup during bone drilling: A brief review on the comparison of thermocouple and infrared thermography

Md Ashequl Islam[1], Nur Saifullah Kamarrudin[1,2], Ruslizam Daud[1], Ishak Ibrahim[1,2], Anas Rahman[1,2] and Fauziah Mat[1]

[1]Faculty of Mechanical Engineering Technology, Universiti Malaysia Perlis, 02600 Arau, Perlis, Malaysia.

[2]Centre of Excellence Automotive & Motorsports (MoTECH), Pauh Putra Campus, Universiti Malaysia Perlis (UniMAP), 02600 Arau, Perlis, Malaysia.

Email: ashequl.unimap@gmail.com

**Abstract.** Predicting thermal response in orthopedic surgery or dental implantation remains a significant challenge. This study aims to find an effective approach for measuring temperature elevation during a bone drilling experiment by analyzing the existing methods. Traditionally thermocouple has frequently been used to predict the bone temperature in the drilling process. However, several experimental studies demonstrate that the invasive method using thermocouple is impractical in medical conditions and preferred the thermal infrared (IR) camera as a non-invasive method. This work proposes a simplified experimental model that uses the thermocouple to determine temperature rise coupled with the thermal image source approach. Furthermore, our new method provides a significant opportunity to calibrate the thermal IR camera by finding out the undetected heat elevation in a workpiece depth.

## 1. Introduction
Temperature rise is a common bone drilling problem in orthopedic and dental surgery, which is an essential factor that leads to bone cell death, known as Osteonecrosis. Excessive heat is generated during the drilling process, resulting in implant failure or screw loosening. Heat generation during bone drilling operations is a severe challenge in orthopedic and dental surgery. Monitoring temperature is crucial to determine the precise result of the bone drilling parameters [1]. Methods used to record temperature elevation in real-time include either invasive measurement by thermocouple instruments [2] or non-invasive estimation by infrared (IR) thermography. Thermocouples have been used for many years and are the most common tool for calculating temperature differences in vivo and ex vivo experiments [3,4]. However, thermocouples outputs are typically governed by many programs that may render future comparisons somewhat uncertain. The thermographic camera has recently increased to determine bone temperature during drilling due to its ease of temperature assessment at any appropriate positions [5]. This technique is limited for medical use due to the high cost, and the camera cannot be used in combination with irrigation systems. Few studies have compared both techniques, finding it difficult to know which of them is ideally used during bone drilling Surgery. Therefore, the present study aims to compare thermocouples and IR thermographic temperature elevation measurement systems and proposes an appropriate technique.

## 2. Invasive Method using Thermocouple

A thermocouple is a sensor consisting of two different conductors (usually metal alloys) to measure temperature. As the temperature varies between the ends of both conductors, there is a relative difference in the voltage between the conductors and the range of thermocouples available on the market [6]. The difference between them depends on the composition of the metals that determine the temperature range and the sensitivity. Low-sensitivity thermocouples (b, r, and s types) have low resolution, and high-sensitivity thermocouples (e, j, k, and t types) have high resolution. Thermocouples are often designed for a higher and lower temperature range.

Figure 1 presents an example of an experimental setup in which a thermocouple is inserted into the bone sample depth to the closest distance of the drill side wall. The embedded thermocouple is connected to a data logger to provide temperature reading during bone drilling.

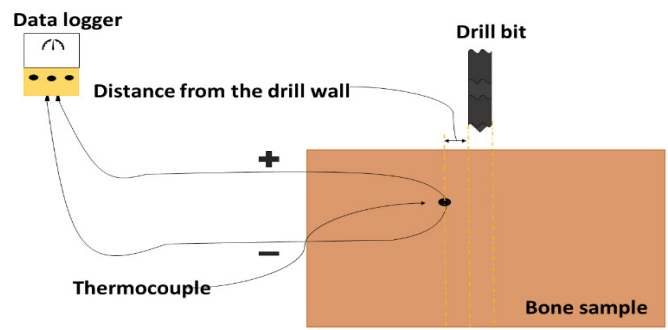

**Figure 1.** Process of Bone Drilling and Temperature Measurement using Thermocouple.

The use of thermocouples in medicine was first reported in dentistry by Horch and Keiditsch [7]. Thermocouples also are mounted inside the drill bit with the drill bar glued around the drill barrel, allowing the temperature to be measured by rotating the instrument [8]. Other strategies are inserting several thermocouples at multiple locations on the surface or bone tissue depth. Thermocouples are usually positioned at an increasing radial distance of 0.5 mm, 1 mm, and 3 mm from the drilling site, which is ideal for comparing the effects [4]. However, various published experiments are using different thermocouples at different lengths, bone models and densities. For easier reading, some articles published between 2015 and 2020 using thermocouples to assess the amount of heat generated in the experimental design shown in Table 1.

**Table 1.** Some Experimental Studies of Heat Measured using Thermocouples.

| Article | Bone type | Sensors | No. of thermocouple | Distance from the hole wall (mm) | Depth (mm) |
|---|---|---|---|---|---|
| Palmisano et al. (2015) [9] | Cadaveric tibia | K type | 4 | 3 | 2 |
| Alam K et al. (2015) [10] | Cortical femur | K type | 1 | 1 | 4 |
| Yali Hou et al. (2015) [11] | Ox thighbone | K type | 1 | 0.5 | 6 |
| Delgado-Ruiz et al. (2016) [8] | Bovine | T type | 2 | 0.5 | 3, 10 |
| Vishal Gupta et al. (2016) [12] | Pig | K type | 4 | 0.5, 1, 1.5, 2 | - |
| Sönke harder et al. (2018) [13] | Bovine | T type | 3 | 0.5 | 4, 8, 10 |
| Jianbo Sui et al. (2019) [14] | Cow femur | T type | 3 | 0.5 | 0, 1, 2.5 |

The literature suggests that the thermocouples selected for bone models are standard of type K. Most of them used two or more thermocouples and are imposed on a bone specimen at radial distances of 0.5 mm and 1 mm. A few of the experiments have been performed over longer distances. Although thermocouple technology is well known, research on its usage is still not reliable. The authors have illustrated other concerns associated with the use of thermocouples. Concerns have been raised about the ability of thermocouple to sense only spot temperature. The technology does not allow creating an overall thermal profile or heat calculation that has leaked through the holes. Another disadvantage is the technical challenge often faced during the insertion of the components very close to the drill. The thermocouple calculation is not a reasonable approach due to low thermal conductivity and inconsistent bone properties. Besides, a significant number of pilot holes must be prepared for thermocouple injection during tests. IR technology has been employed to overcome these difficulties.

**3. Non-Invasive Method Using Infrared (IR) Thermography**
Infrared thermography is the process of detecting radiation (heat) emitted by an object, converting it to temperature, and displaying an image of the temperature distribution using a thermal imager. Since this energy depends on the temperature of the body being studied, a precise determination of the temperature may be made. The acquired signals are used to create a two-dimensional (2D) thermal image. A 3D surface can be reconstructed by integrating two or more cameras.

Figure 2 illustrates the measuring temperature rise with a thermal infrared (IR) camera near the drilling site. The IR thermal camera connects with a computer in which the raw data of temperature distribution could be stored in real-time. The IR thermal camera would make the process of monitoring temperature elevation convenient in a bone drilling experiment. The thermal IR camera should be placed near the drilling site to capture the thermal distribution more accurately.

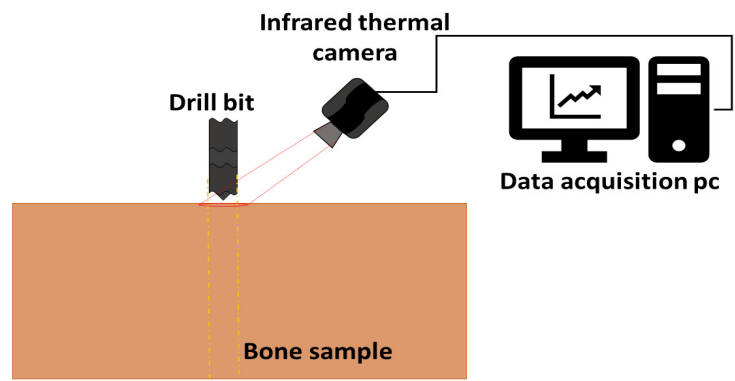

**Figure 2.** Process of Bone Drilling and Temperature Measurement using IR Camera.

The IR thermography technique is a non-contact temperature measuring method widely used in research related to biomedical engineering. As described above, some scholars have recently proposed using thermography as a more reliable tool than thermocouples because of their overall heat assessment. Furthermore, this procedure does not require direct interaction and is thus more practical during in vivo or actual surgical trials [8]. The total heat map created provides the spatial distribution of the heat and the affected area. Many experiments have used thermography techniques to measure the amount of heat produced in real-time, as shown in table 2.

Table 2. Some Experimental Studies of Heat Measured using IR Thermography.

| Article | Bone type | Temperature measurement system | No. of cameras | Distance from the hole wall (m) |
|---|---|---|---|---|
| M. Ghazali et al. (2018) [15] | Pig femur | Therma Cam Flir E300 | 1 | 2 |
| Sönke harder et al. (2018) [13] | Bovine | Fluke Ti20, Fluke Germany | 1 | 0.5 |
| Qingchun Zheng et al. (2018) [16] | Femur | Fluke Tix660 60 Hz | 1 | 0.5 |
| Seifollah Gholampour et al. (2019) [5] | Femur | Flir C2 | 1 | 0.5 |
| Yue Zhang et al. (2019) [3] | Pig femur | TVS-500EX NEC, Japan | 1 | 0.3 |

All experiments use only one camera. However, as mentioned above, using two or more cameras will recreate 3D images but still reflect the surface temperature, even though it is a 3D surface. Despite all its advantages, IR thermography still has some drawbacks, either due to lack of evidence on bone tissues thermal properties or the technique itself. This measurement technique can only detect the surface temperature, which renders it impossible to determine the extent of the necrotic region's depth. All strategies have drawbacks. IR thermography can be used concurrently with thermocouples to achieve an overall understanding of the temperature profile.

**4. Thermal IR Camera Calibration Design Method**
It is essential to find a relation between temperature elevation measurement using a thermographic camera and increased depth in the bone. As shown in figure 3, the recommended experimental design includes a thermocouple. The temperature that the thermal IR camera cannot detect would be possible to calculate by comparing it with a thermocouple placed in depth. The non-identified temperature will be calibrated in the thermal camera with a relevant depth of the drill.

A relation between detected values from the thermocouple and IR camera needs to be measure first. Therefore, a calibration between known temperature from thermocouple will be plotted with captured temperature in IR camera. Calibration will be repeated with several known temperatures at thermocouples, e.g., from 30°C - 60°C with increasing every 5°C.

The next step is to find a relation between the thermocouple and thermal IR camera measured temperature on the surface (figure 3a) and osteons depth during drilling (figure 3b). Figure 3(a) shows that the Thermocouple 1 and IR thermal camera measures the surface temperature and are assumed to be in the same horizontal line during bone drilling. However, in the experiment in figure 3(b), the Thermocouple 2 in $x$ mm depth was inserted into the bone sample. It is expected a temperature difference between Thermocouple 1 and 2, and the relationship will be investigated further.

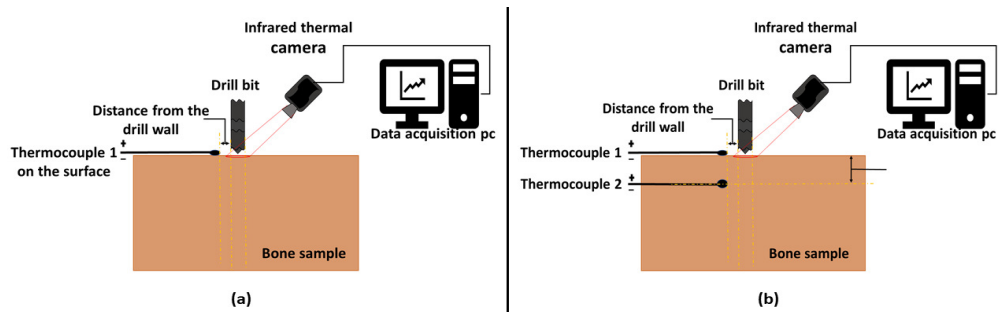

**Figure 3.** Temperature Measurement on the Surface(a) and Temperature in-depth (b) of a Bone Sample using Thermocouple and Thermal IR Camera.

## 5. Conclusion

Infrared thermography usage is preferable due to its practicality in contrast to thermocouples. Another downside to the use of thermocouples is that most versions are not appropriate for in vivo experiments. Thermocouple essentially reduces the scope for further advancement of in vivo experiments which could be of potential value in the future. Although IR thermography records significantly lower temperatures than thermocouples, this technique helped identify heat distribution and ease of setting up in required positions during drilling. So, the calibrated thermal IR camera would replace the embedded thermocouple heat measurement technique in an osteotomy during the preparation and insertion of implants.